# Temperature and Pressure Dependent Luminescence Mechanism of Zinc-Blende Structured ZnS:Mn Nanophosphor under UV and X-ray Excitations

A.K. Somakumar [a], Y. Zhydachevskyy [a,b], D. Wlodarczyk [a], S.S Haider [a], J. Barzowska [c], K.R Bindu [d], Y. K. Edathumkandy [a], Tatiana Zajarniuk [a], A. Szewczyk [a], S. Narayanan [a], A. Lysak [a,b], H. Przybylińska [a], E.I Anila [e], A. Suchocki [a*]

[a] Institute of Physics, Polish Academy of Sciences, Al. Lotnikow 32/46, 02-668, Warsaw, Poland

[b] Berdyansk State Pedagogical University, Shmidta Str. 4, Berdiansk 71100, Ukraine

[c] Institute of Experimental Physics, Faculty of Mathematics, Physics and Informatics, University of Gdansk, ul. Wita Stwosza 57, 80-952 Gdańsk, Poland

[d] Sree Sankara Vidyapeetom College, Valayanchirangara, Kerala, 683556, India

[e] CHRIST (Deemed to be University), Bengaluru, Karnataka, 560029, India

## Abstract

A comprehensive photoluminescence and mechanoluminescence analysis of ZnS:Mn$^{2+}$ nano-phosphor with zinc blende structure is presented. The sample containing quantum dot-sized nanocrystallites were synthesized by the chemical precipitation method and shows excellent orange luminescence at ambient conditions related to the $^4T_1 \rightarrow {}^6A_1$ transition. The sample shows stable and identical luminescence behavior under both UV and X-ray excitation at ambient conditions and also shows excellent self-powered mechanoluminescence properties. The pressure and temperature-induced luminescence mechanism of the phosphor is also established. The shift of the $^4T_1 \rightarrow {}^6A_1$ luminescence band of Mn$^{2+}$ with both pressure and temperature and the luminescence mechanism is explained via the d$^5$ Tanabe Sugano diagram. The broad luminescence band of $^4T_1 \rightarrow {}^6A_1$ transition shifts from visible to near-infrared range at a rate of -35.8 meV/GPa with the increase of pressure and it is subsequently quenched completely at a pressure of 16.41 GPa due to a reversible phase transition from zinc blende (F$\bar{4}$3m) to rocksalt (Fm$\bar{3}$m) phase. The high-pressure and temperature-dependent decay kinetics measurements of the sample luminescence are also reported.

**Keywords:** Zinc Sulphide, Zinc blende, Manganese doping, Tanabe-Sugano diagram, High-pressure luminescence, Decay kinetics, Mechano-luminescence,

# 1. Introduction

Zinc Sulphide (ZnS) is one of the oldest and most widely used luminescent host materials for lighting and scintillating applications[1]. It is also well known for its application as a scintillating screen in the famous Rutherford Alpha scattering experiment [2]. ZnS-based pure, doped, and composite detectors and scintillators continue to be widely explored[3,4]. Doping ZnS with various activators like Al, Ag, Cu, and Mn is particularly interesting because of features like high-intensity broad emissions in multiple colors, sensitivity towards various types of radiations, and relatively simple growth and synthesis procedures [5–8].

In that, bulk and nanoparticle-sized ZnS samples doped with manganese (Mn) are well-examined by researchers due to their various luminescence features in the visible regime. However, the nano-phosphor of ZnS:Mn samples with quantum dot-sized particles are particularly interesting because their luminescence-based features are enhanced due to the quantum confinement effects[9–11]. Generally, Mn-doped ZnS is not considered a good scintillator compared to silver-doped ZnS due to its millisecond decay times and after-glow effects compared to the ZnS:Ag, however, it still possesses some significant X-ray and UV radiation detection capabilities[2]. Many of the papers regarding high-pressure photoluminescence studies of ZnS:Mn quantum dots with zinc blende structure did not go up to the phase transition to the rock-salt phase [12–14]. However, this is particularly important to know the phase changes happening for the ZnS material used in the luminescence and radiation detectors working under extreme conditions like high pressure and varying temperatures. It is particularly interesting in the various outer space applications of ZnS:Mn material, which usually happens in extreme conditions[15,16]. Compared to many other space-based detectors, the ZnS-based ones can also sense impacts from objects without any external illumination, this advantage is particularly important when deploying radiation monitoring sensors in outer space to sense radiations. Moreover, these detectors can detect micrometeoroid and space debris impacts on their surfaces [17–19]. However many of the powder-based detectors do not show both radiation sensing and impact-induced mechanoluminescence together without external illumination. Also, in the past researchers explored the possible application of ZnS:$Mn^{2+}$ to detect foreign body impacts on outer space vehicles and as a lightweight phosphor-based health monitoring sensor suite for the same spacecraft [20,21].

Generally, at ambient pressure, the ZnS phosphor materials are mainly available in two different crystalline forms zinc blende (cubic) and wurtzite (hexagonal), in which the former is the most commonly available. For the present work, we chose ZnS:$Mn^{2+}$ samples with zinc blende structure containing extremely fine particle sizes, for the low-temperature, high-pressure, mechanoluminescence studies, and also tested its luminescence decay behavior as a function of temperature and pressure. Also, we did a detailed analysis of previous high-pressure measurements of the ZnS:Mn samples with zinc blende structure.

## 2. Experimental

### 2.1. Materials and synthesis

The manganese-doped ZnS nanophosphor was synthesized by the same chemical precipitation method previously reported in the literature[10]. 25 milliliters of $Zn(CH_3COO)_2$, $MnCl_2$, and $Na_2S$ water solutions were utilized to prepare initially for the synthesis of $Mn^{2+}$ doped ZnS nanoparticles. One molar $Na_2S$ solution was added drop by drop to the mixture of 1 M zinc acetate $Zn(CH_3COO)_2$ and 0.02 M $MnCl_2$ while being continuously stirred using a magnetic stirrer. The mixture was swirled while maintaining a steady temperature for 20 minutes. After filtering the resulting white colloidal suspension, the filtrate was cleaned with deionized water and allowed to dry for a day at 70 °C. As synthesized white-colored powder of ZnS phosphor nominally doped with 2 at.% Mn was used for further analysis.

### 2.2. Experimental techniques

The x-ray diffraction (XRD) experiment was performed with a BRUKER D2PHASER using Cu Kα radiation operating at 30 kV and 10 mA. The XRD pattern was collected with a scan step of 0.02° and an acquisition time of 1s per step. The analysis was performed using DIFFRAC.EVA V4.1 software from BRUKER and ICDD PDF-4 database (release 2021). The morphology analysis of the prepared sample was done with a Hitachi SU-70 Scanning Electron Microscope (SEM). The ambient temperature Raman spectrum of the ZnS:Mn sample was recorded with a Monovista CRS+ Raman spectrometer from S&I Gmbh equipped with a nitrogen-cooled CCD detector (-125 °C) and a 0.75 m monochromator from Acton Princeton Ltd. with a holographic grating of 2400 grooves/mm. A 785 nm Torsana StarBright L series laser was used to illuminate the sample. Photoluminescence emission, excitation, and decay studies were done on the Horiba Fluorolog-3 modular spectrofluorimeter with a 450 W xenon lamp source. A Janis continuous-flow liquid helium cryostat coupled with a Lake Shore 331 temperature controller was used for the temperature-dependent studies. The room temperature radio-luminescence emission spectrum was recorded using the same spectrofluorimeter and a Hamamatsu L9181-02 model continuous Microfocus X-ray source powered with a voltage of 130 kV and 300 µA current. The sample was kept in an aluminum crucible and placed roughly 57mm away from the X-ray source window for the measurement. The high-pressure luminescence measurements were performed using a miniature Diamond Anvil Cell (DAC) supplied by easyLab with a culet diameter of 0.45 µm. As a pressure-transmitting medium a mixture of methanol and ethanol in a volume ratio of 5:1 was selected. To collect the luminescent emissions from the sample, a backscattering geometry was utilized. For the data collection, a Horiba Jobin Yvon Triax320 monochromator equipped with a Spectrum One liquid-nitrogen-cooled CCD camera was employed. The sample was excited using a 325 nm laser line generated by a 20 mW He−Cd laser. This excitation wavelength was specifically chosen because it is very close to the broad charge transfer band. The high-pressure decay analysis was performed with the same diamond anvil cell method and a pulsed laser excitation from NT342/3 series tunable optical parametric oscillator (OPO) equipment with an Nd:YAG laser from EKSPLA. The emitted photons were counted using a Micro Photon Devices (MPD), PDM series photon counting module attached to a Princeton Instruments Acton Spectra Pro SP-2500 monochromator. The impact-induced mechano-luminescence measurement of the prepared phosphor was done at ambient pressure and temperature conditions on a custom-built equipment described in the literature[22]. The innovative setup consists of a CM.122 airsoft electric gun, speedometers, target cylinder, and condenser coupled with a photomultiplier and a digital oscilloscope using optical fiber. The room temperature EPR spectra of zinc blende and wurtzite samples were recorded using a Bruker EMX spectrometer operating at a frequency of 9.5 GHz (X band). The samples were kept inside a quartz tube and placed inside the microwave cavity during the measurement and the magnetic field from the EPR spectrometer was precisely limited to the horizontal plane while the microwave field rf had a vertical polarization. Field-dependent magnetization measurements at temperatures of 5 K and 300 K, in magnetic fields up to 90 kOe, were carried out using a sample vibrating magnetometer of the Physical Property Measurement System (PPMS-9T) from Quantum Design.

## 3. Results

### 3.1 Structure and morphology

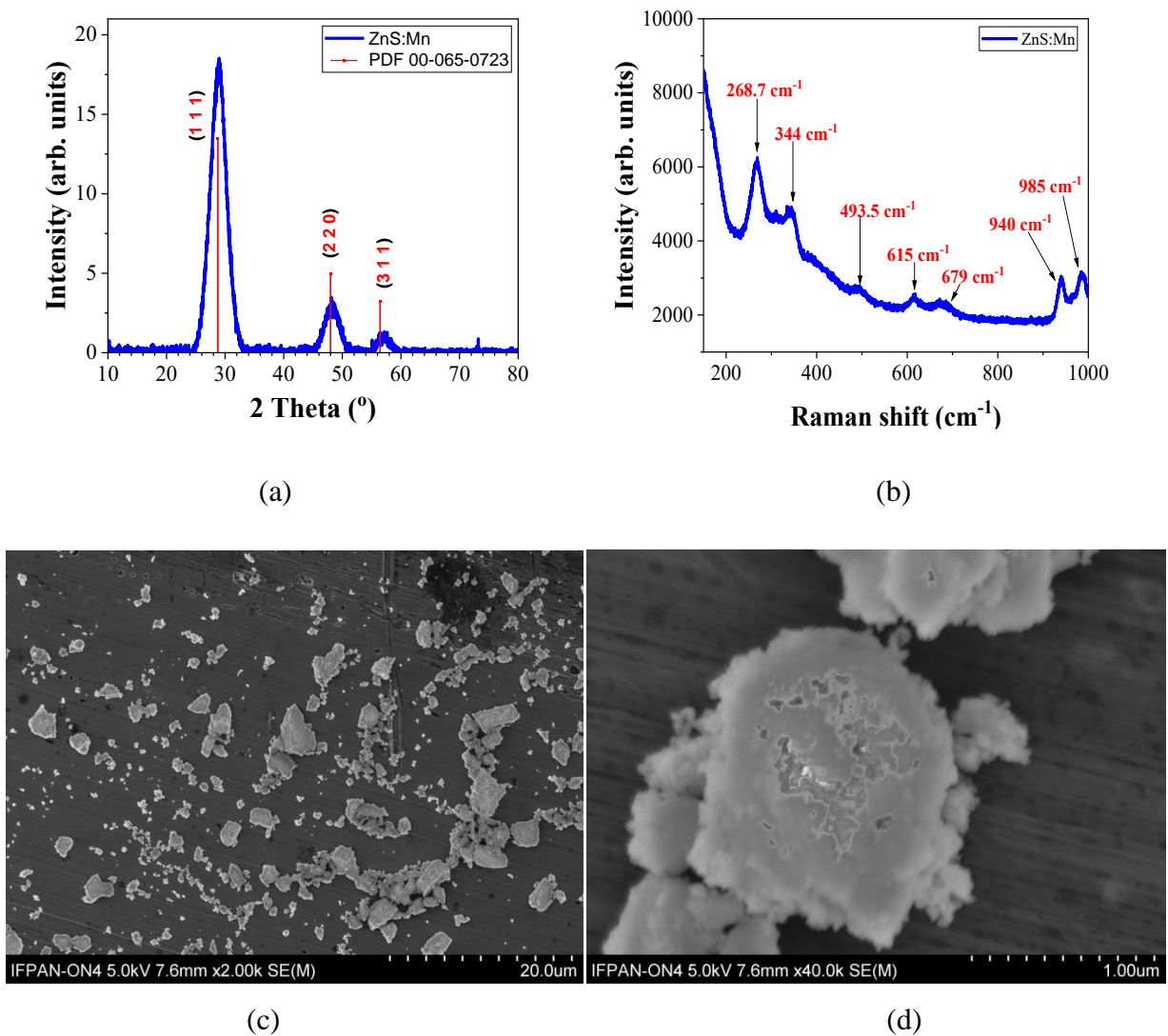

Figure 1. (a) XRD pattern (b) Raman spectra and (c) SEM image of prepared ZnS:Mn sample (d) enhanced SEM image.

The X-ray powder diffraction data of the ZnS:Mn sample in Figure 1(a) shows that the prepared sample has broad peaks at 2θ values of 28.9°, 48.2° and 57.2° and those peaks perfectly match with the crystalline planes (111), (220), and (311) reported for the zinc sulfide sample with a cubic zinc blende structure in ICDD PDF card number 00-065-0723 with a space group of $F\bar{4}3m$. Apart from these three prominent peaks, we were not able to find patterns related to any other phases or impurities in the spectrum. That points towards the formation of pure zinc blende phase of the prepared ZnS:Mn sample and also the broadness of the XRD peaks confirms the strong quantum confinement effects present in the prepared nanoparticles. In the zinc blende structure, each zinc ($Zn^{2+}$) ion tetrahedrally coordinated with four other sulfur ($S^{2-}$) ions and vice versa. After doping with manganese ($Mn^{2+}$) the $Mn^{2+}$ ions occupy the $Zn^{2+}$ ions positions in the lattice. However, the small difference in ionic radii between $Mn^{2+}$ (0.066 nm) and $Zn^{2+}$ (0.06 nm) creates a lattice misalignment, which enhances the formation of crystallites with quantum dot size due to high strain energy[23]. The size of the nanocrystallites in the sample was estimated with the Debay-Scherrer method:

$$D_{hkl} = \frac{K\lambda}{(\beta cos\theta)} \qquad (1)$$

were the parameters K= 0.89 and apparatus broadening equal to 0.05, and obtained a crystallite size of 2.38 nm. So the prepared phosphor nanoparticle has a quantum dot size that falls well within the range of the Bohr exciton radius of ZnS, which is close to 2.5nm[24]. The Raman spectrum of ZnS:Mn particles in Figure 1(b) have been recorded at ambient temperature and pressure. There are two prominent broad peaks observed at 268.7 $cm^{-1}$ and 344 $cm^{-1}$ initially, which are related to the first-order transverse-optical (TO) and longitudinal-optical (LO) phonons of the ZnS respectively[25,26]. The observed energy shift at 615 $cm^{-1}$ and 679 $cm^{-1}$ is related to the second-order Raman TO and LO phonons respectively[27,28]. However, these second-order peaks are comparatively very weak in this sample. The presence of these Raman peaks at respective positions also confirms the cubic nature of the prepared ZnS:$Mn^{2+}$ phosphor. The broadening of the first-order TO and LO bands is probably related to the quantum confinement effect of the ZnS:$Mn^{2+}$ quantum dot mentioned previously. The surface morphology analysis of the prepared sample was done with a scanning electron microscope (SEM). Figure 1(c) shows that the powder sample was composed of agglomerated formations of nanocrystallites with various shapes. The enhanced SEM image in Figure 1(d) depicts the rough surfaces of the agglomerated nanocrystalline formations.

## 3.2 Ambient pressure luminescence studies

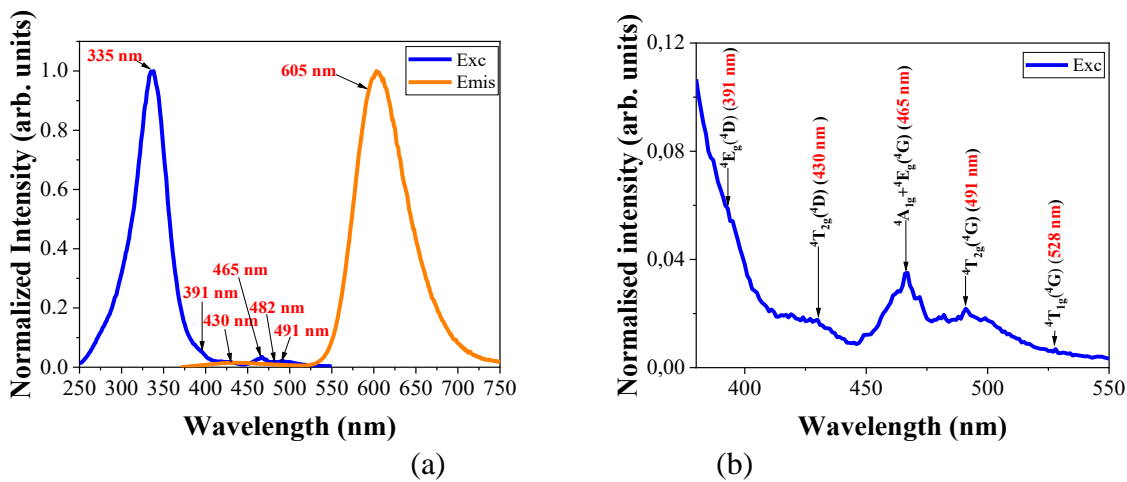

Figure 2.(a) Photoluminescence emission spectra with 335 nm UV excitation and excitation spectra recorded for 605 nm emission peak (b) enlarged view of the d-d transition bands in the excitation spectra.

The luminescence study of the 2% Mn-doped ZnS sample was initially done under ambient temperature and pressure conditions. According to Bhargava et al., the bandgap energy of ZnS nanocrystals increases from 3.66 eV to higher energies with the decrease in size due to the quantum confinement effect[11,29]. The sample shows intense photoluminescence at ambient conditions. Figure 2(a) shows the normalized photoluminescence excitation and emission spectra of the sample. The figure depicts the excitation spectrum which consists of a relatively intense broad band centered around 335 nm and many other small peaks related to other transitions, however, these smaller peaks are attributed to the d-d transitions between the energy levels of the $Mn^{2+}$ ion. Figure 2(b) shows the enlarged view of these d-d transition bands. The experimentally observed transitions are listed in the following Table1.

Table 1: Position of lowest quartet excited states of $Mn^{2+}$ ions in ZnS recorded for 605 nm emission.

| Terminating state | Wavelength and energy of the PLE peak |
|---|---|
| $^4T_{1g}$ ($^4G$) | 528 nm (18939 cm$^{-1}$) |
| $^4T_{2g}$ ($^4G$) | 491 nm (20367 cm$^{-1}$) |
| $^4A_{1g}$ + $^4E_g$ ($^4G$) | 465 nm (21505 cm$^{-1}$) |
| $^4T_{2g}$ ($^4D$) | 430 nm (23256 cm$^{-1}$) |
| $^4E_g$ ($^4D$) | 391 nm (25575 cm$^{-1}$) |
| C.T | 335 nm (29851 cm$^{-1}$) |

The emission spectrum consists of two different bands: one broad blue luminescence band around 435 nm and an orange luminescence band around 605 nm. The blue band is probably related to the defects in the host material and the orange band is due to the $^4T_1 \rightarrow ^6A_1$ transition of the $Mn^{2+}$ dopant, replacing $Zn^{2+}$ ions. As per some previous studies when the Mn doping percentage increases in the matrics, the $^4T_1 \rightarrow ^6A_1$ emission-related band moves towards lower energies. Moreover, the intensity of this band grows with the increase of Mn doping concentration[23]. In addition to that, the position of this main luminescence band slightly varies with the change of the nanoparticle size [30]. A relatively less intense blue luminescent band (435nm) compared to the highly intense orange emission at 605nm is shown in Figure 2 (a). It confirms the $Mn^{2+}$ ions are perfectly doped inside the ZnS host lattice. The optical properties of the $Mn^{2+}$ ions in ZnS can be explained by the $d^5$ state of the Tanabe Sugano diagram (described in the discussion part).

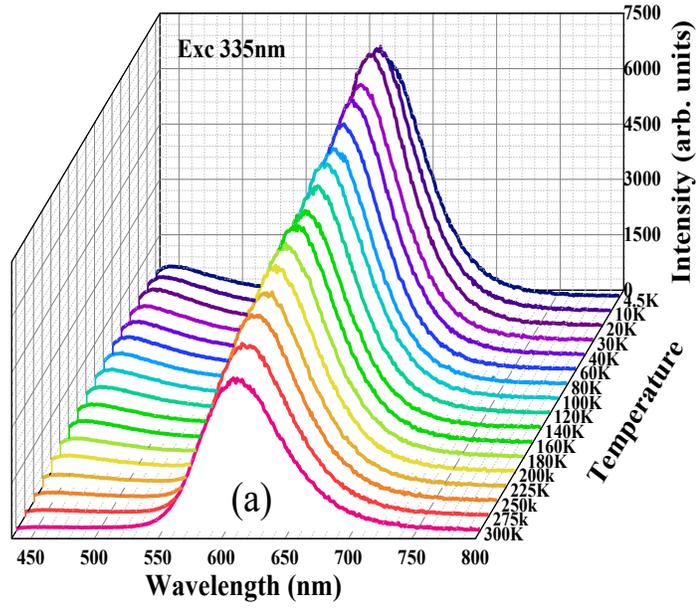

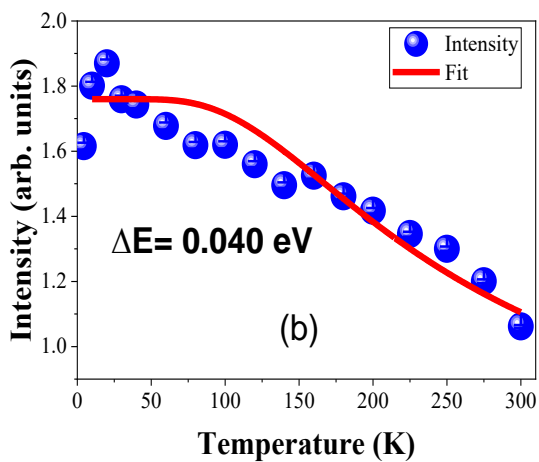
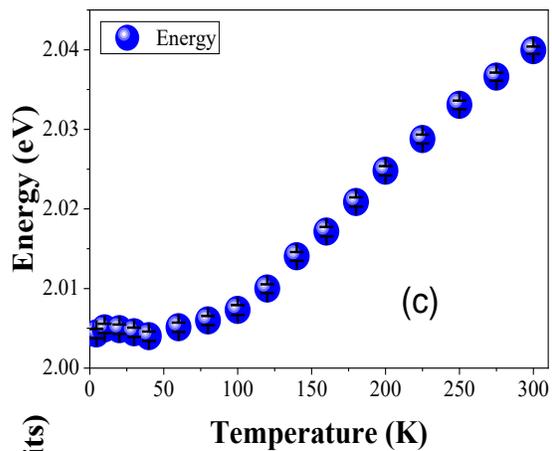
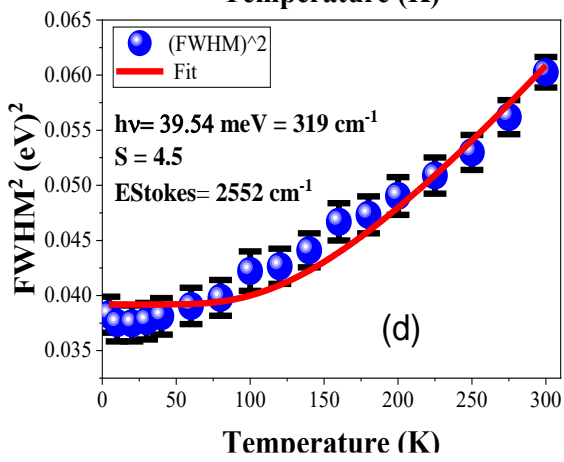
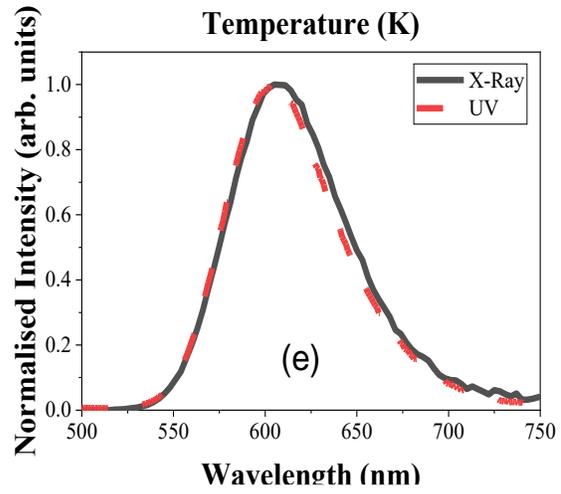

Figure3.(a) Temperature dependence of $Mn^{2+}$ luminescence spectra (b) integrated emission intensity versus temperature ( the red line shows the fit) (c) the energy of $Mn^{2+}$ luminescence band versus temperature and (d) full-width at half maximum of the emission peak versus temperature( the fit marked as red) (e) comparison of X-ray excited radio-luminescence spectra and UV excited luminescence spectra of ZnS: $Mn^{2+}$ nanophosphor.

Figure 3(a) shows the luminescence spectra of $Mn^{2+}$ doped ZnS as a function of temperature with 335 nm xenon lamp excitation. The intensity of the main $^4T_1 \rightarrow {}^6A_1$ luminescent band at 605 nm decreases with an increase in temperature from 4.5 to 300 K. Also, this band shifts towards higher energies with the increase in temperature (see Figure 3 (c)). The luminescence band near the blue region also shows a thermal quenching with the increase in temperature, however, the luminescence emission quenching of the blue band is comparatively faster than the orange luminescence band at 605 nm. In earlier comparative studies between Mn-doped ZnS samples and those doped with Copper (Cu) and Europium (Eu), it was observed that the emission from manganese-doped ZnS has a significantly slower quenching rate with the increase of temperature in contrast to the other dopants[31]. Figure 3(b) shows the integrated emission intensity (arb. units) versus temperature (T) graph of $Mn^{2+}$ luminescence in ZnS. The graph was fitted with the Arrhenius equation:

$$I(T) = \frac{I_o}{1+Ae^{-\frac{\Delta E}{K_B T}}} \quad (1)$$

where $I_o$ is the initial luminescence intensity of the $Mn^{2+}$ luminescence band, $\Delta E$ is the activation energy, and $K_B$ represents the Boltzmann constant. The fit of equation (1) (marked with red) in Figure 3(b) shows that the intensity of the prominent $Mn^{2+}$ band decreases with the increase of temperature from 4.5 K to 300 K and it also indicates the thermally activated energy transfer process with an activation energy of about 0.040 eV (40 meV). Figure 3(c) & (d) shows the energy and FWHM versus temperature graphs, it depicts that the $Mn^{2+}$ emission band blue-shifted to a shorter wavelength (or higher energies) with an increase in temperature. This type of behavior is non-typical and may testify to unusual thermal properties of ZnS, such as theoretically postulated a certain time ago negative thermal expansion (at least at low temperatures)[32]. Simultaneously the FWHM of the main luminescence band was increased with the increase in temperature. The FWHM versus temperature dependence is fitted with the following equation:

$$\text{FWHM} = 2.36\sqrt{S}\hbar\Omega\sqrt{\coth\left(\frac{\hbar\Omega}{2kT}\right)} \qquad (2)$$

where S is the Huang-Rhys factor, and its value determines the strength of the electron-phonon coupling in both the excited and ground state of the luminescence center and $\hbar\Omega$ is the effective phonon energy. The theoretical calculations show that the S value in the present experiment is 4.5 and it is relatively closer to the strong coupling regime (it is $S \geq 5$) [33]. Previous studies on the ZnS:$Mn^{2+}$ sample have also shown that the S value is greatly influenced by the size of the nanoparticles[30]. The effective phonon energy $\hbar\Omega$ is around 39.54 meV (or 319 $cm^{-1}$), which is very close to the first-order LO phonon frequency measured at around 42.7 meV (or 344 $cm^{-1}$) in the Raman spectra. Both are related by the following equation:

$$E_{Stokes} = (2S - 1)\hbar\Omega \qquad (3)$$

According to equation (3) the theoretical Stokes shift value is equal to $E_{Stokes}$= 2552 $cm^{-1}$. This theoretically obtained Stokes shift value added to the emission band centered at 605 nm, yields the expected absorption peak of $^6A_{1g} \rightarrow {}^4T_{1g}$ ($^4G$) transition around 524 nm, which is very close to the experimentally observed value presented in Table1.

### 3.3 Radioluminescence

The sample was investigated further with another type of excitation to check the $Mn^{2+}$ luminescence consistency with other types of radiations, for that the sample was excited with a continuous microfocus X-ray source. Figure 3(e) shows the X-ray excited radio-luminescence spectrum of ZnS:$Mn^{2+}$. Depicted luminescence curve is very similar to the spectrum obtained in Figure 2(a) with UV excitation, consisting of both blue and orange luminescence bands. However, the defect-related emission band in the blue region was negligibly weak with the X-ray excitation. Figure 3(e) also shows the comparison of the main luminescence band related to the $^4T_1 \rightarrow {}^6A_1$ transition of $Mn^{2+}$ with both X-ray and UV excitations. It shows that both UV and X-ray excited emissions of $Mn^{2+}$ are identical to each other at ambient conditions and the value of FWHM of both luminescence bands are also very close and comparable, for UV it is 70.81 nm and for X-ray, it is about 72.63 nm.

## 3.4. Impact-Induced Mechano luminescence & EPR Analysis

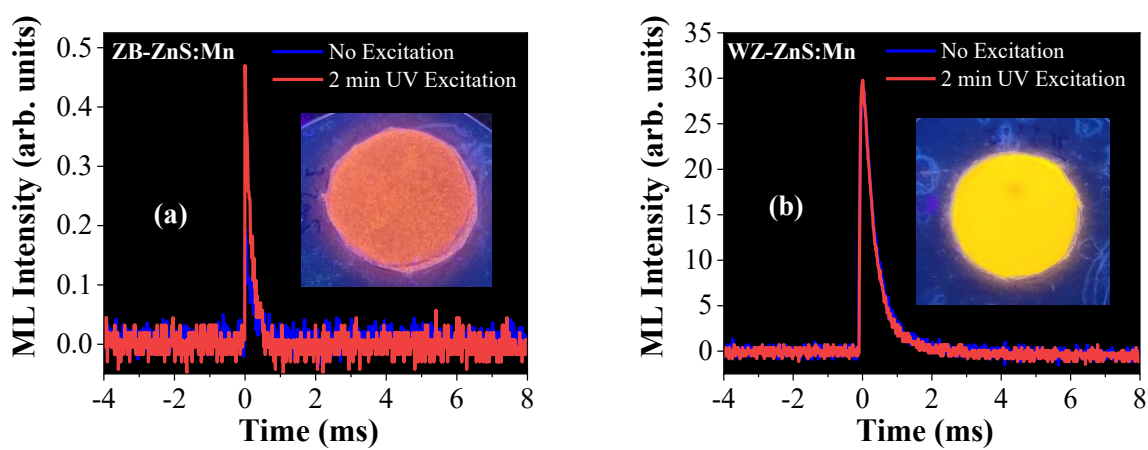

Figure 4. I-ML intensity versus time plot of ZnS:Mn samples ((a) zinc blende and (b) wurtzite) without any earlier irradiation and with earlier UV irradiation for 2 min (inset images show the respective sample films prepared for the I-ML experiments under UV light)

Figure 4(a) & (b) shows the impact-induced mechanoluminescence (I-ML) response of the ZnS:Mn samples with respect to the time, collected when the sample was hit by a bullet from an air-soft electric gun. The bullet used for the impact studies in both cases had the same size and weight, as well as the same initial kinetic energy. The blue line in Figure 4(a) & (b) shows the I-ML response without any earlier illumination and the red one shows the response curve after initial UV illumination on the sample for 2 minutes. When the bullet hits the sample, a piezoelectric field is created near the activator ions in the hitting area, which lowers the trap depths near activator ions and causes release of electrons from filled-electron traps to the conduction band [34]. This recombination of the electron-hole pair releases some non-radiative energy which excites the $Mn^{2+}$ ions in the lattice and produces (orange in the case of zinc blende or yellow in the case of wurtzite sample, respectively) mechano-luminescence with impact. When it is illuminated with UV more traps are filled with electrons and more recombination processes take place and emit more luminescence compared to others. Moreover, the current zinc blende structured ZnS:Mn sample shows a very good impact-induced mechano-luminescence response even without any external illumination and shows an improved luminescence response when the sample was previously illuminated with a UV excitation. However, Cu or Ag-doped ZnS powders with scintillating properties do not show much mechano-luminescence when compared with $Mn^{2+}$ doped [35]. For a better understanding, we compared the I-ML of our zinc blende sample with a wurtzite type ZnS:Mn sample at the same experimental conditions described previously. In comparison with the zinc blende, the wurtzite ZnS:Mn sample in Figure 4(b) shows a much more intense and slightly broader I-ML response with time. This effect was previously related to the higher piezoelectric coefficient of wurtzite ZnS in comparison with a piezoelectric coefficient of the ZnS with zinc blende structure. However, it did not show any improvement with the earlier illumination with a UV source for 2 minutes compared to the zinc blende. Moreover, the room temperature EPR spectra of both samples were collected at the same experimental conditions shown in Figure S1, which shows the wurtzite sample is nearly 20 times more intense due to the presence of a much higher Mn percentage than the zinc blende sample we synthesized, which is also a reason for the higher luminescence efficiency for the wurtzite sample in the mechanoluminescence spectra in Figure 4 (b).

On the other hand, the mechanoluminescence intensity enhancement for the wurtzite phase over zinc blende was previously reported for a device fabricated with the ZnS material and also they observed ML intensity enhancement while heating the sample due to a temperature-induced phase transition from zinc blende to wurtzite, which also confirms our observations[36]. That means, the ZnS:$Mn^{2+}$ phosphor samples in both phases are ideal for their direct application in self-powered mechanoluminescence devices but in the case of zinc blende structured sample, mechanoluminescence has the potential to be further enhanced by the tuning of $Mn^{2+}$ dopant concentration, temperature, and external illumination power, etc. Also, both samples are promising when these nanoparticles are embedded in polymer materials like PDMS, PVDF, etc in future [37].

However, the observation of higher Mn concentration in the wurtzite sample in EPR data is further confirmed by the comparison of room temperature and low-temperature SQUID measurements of both samples. Figure S2 shows the SQUID results of both zinc blende and wurtzite ZnS:$Mn^{2+}$ samples. The estimation of the ratio of Mn percentage in both samples by the SQUID method also confirmed the result obtained during EPR analysis. Only a small part of Mn was initially introduced during the synthesis of zinc blende quantum dot enters the final material, despite 0.02 M $MnCl_2$ mixture being added to the sample solution during the synthesis. This gives a possibility of an increase in the ML efficiency of ZB type ZnS:Mn if a higher concentration of dopant is introduced to ZB ZnS. Apparently stronger crystal field experienced by $Mn^{2+}$ ions in the ZB sample testify to smaller cation – ion distances in this structure than in the WZ sample. This may be a reason for difficulties in introducing higher concentrations of larger $Mn^{2+}$ than $Zn^{2+}$. Also, the SQUID measurements in Figure S2 show that at ambient temperature both ZnS:$Mn^{2+}$ samples have a paramagnetic-like behavior.

### 3.5. High-pressure luminescence studies

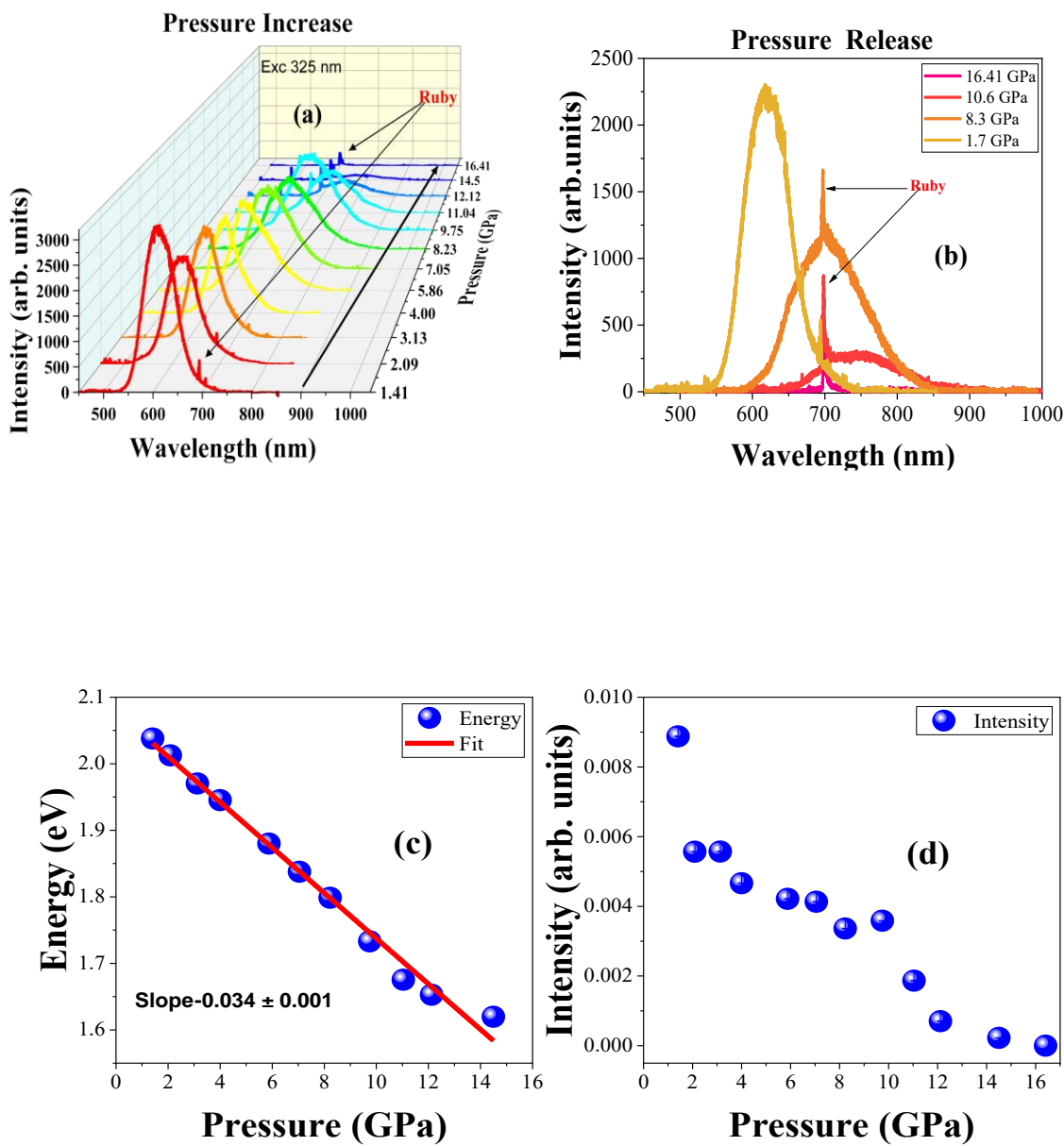

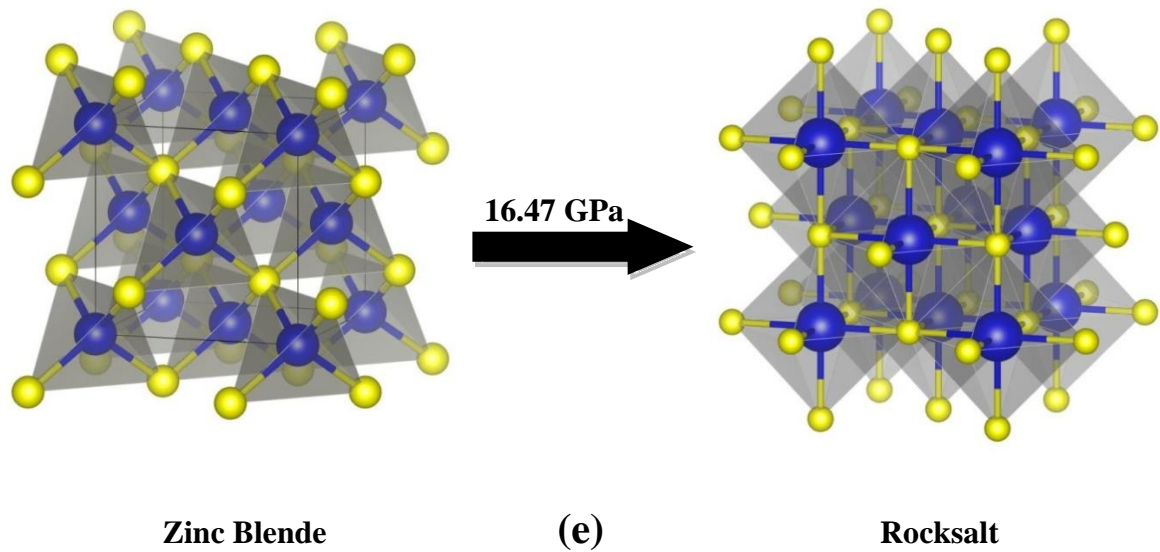

**Zinc Blende** (e) **Rocksalt**

Figure 5. Pressure-dependent luminescence of $Mn^{2+}$ (a) pressure increase (b) pressure release. (c) Energy versus pressure graph and (d) integrated emission intensity versus pressure (e) Phase transition from zinc blende structure to rock salt under high pressure (yellow balls indicate zinc (Zn) and blue balls indicate sulfur (S) atoms, respectively )

Figure 5(a) & (b) shows the pressure-dependent photoluminescence spectra of ZnS:Mn under ambient temperature conditions. Figure 5(a) shows the change in emission spectra during compression and Figure 5(b) shows the emission spectra with the decrease of pressure. Figure 5(a) shows that initially, the spectra were very similar to the one we observed under the ambient condition however at initial pressure it is slightly shifted roughly 610 nm from its original position at 605 nm. It also shows that with the increase of pressure, the $Mn^{2+}$ luminescence gradually shifts up to the NIR region of around 800 nm at high pressure and the luminescence was completely quenched at 16.41 GPa pressure. That means a phase transition occurs around this pressure. Most probably at this pressure the Mn-doped ZnS nanoparticle loses its zincblende structure ($F\bar{4}3m$) and transforms into the rocksalt ($Fm\bar{3}m$) phase. Earlier DFT studies also pointed to the zinc blende to the rocksalt phase transition happening somewhere near 14.7 and 17.6 GPa and it also suggests the zinc blende structure is more compressible than the rock salt one[38,39]. Also, the main luminescence band red-shifted to nearly 775 nm (1.6 eV) during compression, which is in the NIR bio-window region. Figure 8(b) proves that the zinc blende to rock salt phase transition is reversible because the sample luminescence returns to its initial

position and regains its original luminescence intensity after decompression. Previous synchrotron studies of ZnS nanoparticles with a zinc blende structure exhibited a reversible phase transition compared to the ZnS nanoparticle with a wurtzite phase[40]. It is also evident that the grain size of the particle also greatly influences the phase transition pressure of the ZnS nanoparticle [41].

Table.2: Comparison of previous and present high-pressure studies of Zinc Blende structured doped with Mn and undoped ZnS host material.

| Host | Dopant | Sample size | Structure | Study | Max. pressure reached (GPa) | Phase transition (GPa) | Pressure coefficient (meV/GPa) | Year | Reference |
|---|---|---|---|---|---|---|---|---|---|
| ZnS | $Mn^{2+}$ | 1 nm, 3 nm, 3.5 nm, 4.5 nm, 10 nm, Bulk | * | PL | ~7 | No | -31.6 -33.1, -30.9, -31.9, -29 | 2003 | [14] |
| | | 3 nm, 3.5nm, 4.5 nm, 10 nm, bulk | * | PL | ~ 7 | No | -39, -35.7, -33.3, -30.1, -29.4 | 2004 | [13] |
| | | * | * | PL | ~5 | No | * | 1966 | [12] |
| | | 4.5 | Zinc blende | PL | ~7 | No | -33.3 | 2000 | [42] |
| | | * | Zinc blende | PL | ~12 | No | -31 | 1977 | [43] |
| | | Bulk | * | Absorption & PL | ~10 | No | -26.5 | 1988 | [44] |
| | undoped | Bulk 3.5 to 4.5 μm | Zinc blende | Absorption | ~20 | 15 | | 1990 | [45] |
| | | | Zinc blende | X-ray | ~35 | 13.4 | * | 2017 | [46] |
| | | 2.8 | Zinc blende | X-ray | ~20 | 18 | * | 2000 | [40] |
| | | Nanosheets(20-50nm) | Zinc blende | ADXRD[#] | ~33 | 13.1 | * | 2015 | [47] |
| | | Bulk | Zinc blende | Shock | ~ 135 | 15.7 | | 1998 | [48] |

| | | * | Zinc blende | DFT | ~25 | 14.7 | * | 2019 | [38] |
| | | * | Zinc blende | DFT | * | 17.6 | | 2015 | [39] |
| | $Mn^{2+}$ | 2.38 nm | Zinc blende | PL | ~ 17 | 16.41 | -35.8 | Present | |

([#]ADXRD -Angle dispersive x-ray diffraction)

Figure 5(c) shows the linear decrease of energy with the gradual increase of pressure. The pressure coefficient was calculated by the linear fit of the data and it is about -35.8 meV/GPa. It is comparable to the values reported by the previous high-pressure studies of ZnS nanoparticles of similar structure and dimensions shown in Table.2. However compared to the bulk crystal, the ZnS:$Mn^{2+}$ nanoparticle has a slightly higher pressure coefficient due to the higher surface-to-volume ratio[13,14]. The graph in Figure 5(d) depicts the dependence of integrated emission intensity with pressure. It shows the gradual decrease of emission intensity with the increase of pressure. Furthermore, at 16.47 GPa the luminescence intensity was completely quenched. Both graphs in Figure 5(c) & (d) clearly show, that the orange luminescence gradually shifted from the visible (2.037 eV) to the near-infrared region (1.619 eV) and completely disappeared with the next increase of pressure due to phase transformation.

### 3.6. Luminescence decay studies

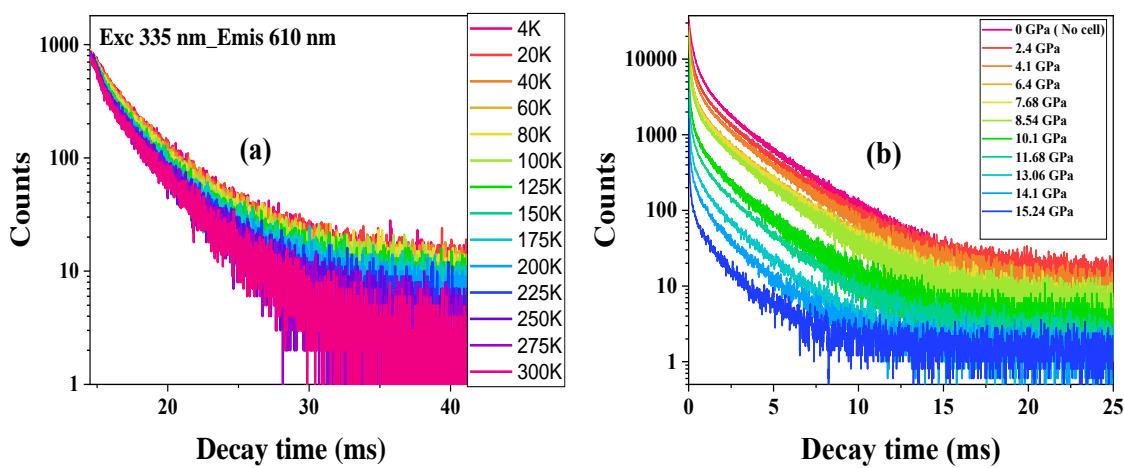

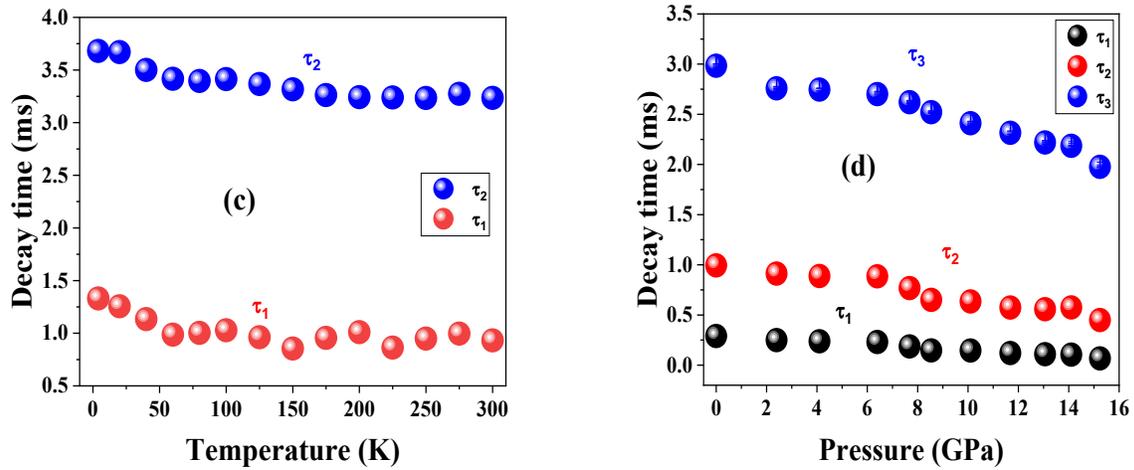

Figure 6. (a) Temperature and (b) pressure-dependent photoluminescence decay kinetics of $Mn^{2+}$ peak in ZnS, decay time versus (c) temperature and (d) pressure graphs.

For a better understanding of the influence of temperature and pressure on the $Mn^{2+}$ luminescence band, the decay analysis of the sample was performed as a function of both temperature and pressure. The temperature-dependent decay measurements were done at different temperatures ranging from 4 K to 300 K shown in Figure 6(a); the sample was excited with a 335 nm xenon lamp excitation and emission was set at the point 610 nm of the main luminescent band related to the $^4T_1 \rightarrow {}^6A_1$ transition. Figure 6(a) shows that the decay time at higher temperatures is a little bit faster than the decay at lower temperatures. The obtained decay curves are fitted with a double exponential equation. Fitted results in Figure 6(c) show that both decay time components $\tau_1$ and $\tau_2$ have a milliseconds lifetime and it decreases with an increase in temperature. Both $\tau_1$ and $\tau_2$ components decrease in a similar manner. Initially, the $\tau_2$ component of the decay time at 4.5 K was around 3.68 ms and it decreased nearly to 3.2 ms at 300 K. The $\tau_1$ was around 1.33 ms at 4.5 K and it decreased nearly to 0.93 ms at 300 K. The milliseconds decay time of the $\tau_2$ component confirms the perfect incorporation of a higher percentage of $Mn^{2+}$ ions inside the ZnS nanoparticles, rather than over the surface. The other $\tau_1$ component indicates the occupation of few percentages of ions closer to the surface.

To know further about the high-pressure phase transition and luminescence quenching behavior of ZnS:Mn sample; the pressure-dependent decay kinetics were also measured. Figure 6 (b) shows the pressure-dependent decay kinetics of $Mn^{2+}$ luminescence. Unlike the temperature-dependent spectra, an additional decay component was also observed during the high-pressure DAC measurement; probably it is due to the use of a pulsed laser excitation source. In an earlier comparison study in the ZnS:$Mn^{2+}$ sample, a similar additional component was observed with pulsed laser excitation than the xenon lamp at ambient conditions [23]. However, the luminescence decay time values for $\tau_1$ and $\tau_2$ at 300 K in Figure 6(c) are close to the $\tau_2$ and $\tau_3$ values of the sample measured at 0 GPa (without DAC) in Figure 6(d). Figure 6(d) also shows that all three lifetime decay components $\tau_1$, $\tau_2$, and $\tau_3$ decrease during compression. The $\tau_3$ component decreased from 2.98 ms to nearly 1.97 ms, the $\tau_2$ component decreased from 0.99 ms to 0.44 ms and the $\tau_1$ component decreased from 0.28 ms to 0.067 ms at a pressure of 15.24 GPa. The $\tau_3$ component shows a larger shift in decay time with compression rather than the two other components. The $Mn^{2+}$ luminescence was completely quenched at the next pressure step due to the induced zinc blende to rocksalt phase transition described earlier, because of that the signal was reduced to background noise. However, earlier pressure-dependent decay analysis of ZnS:Mn sample up to 12 GPa did not show much shift in the main decay component despite an unusual increase in luminescence intensity with high pressure, and also they did not go up to the phase transition point[43].

## 4. Discussion

The observation of pressure and temperature-dependent multi-exponential luminescence decay times in milliseconds indicates the occupation of $Mn^{2+}$ ions in the different places in the ZnS nanocrystal. The longer decay component of nearly 3 ms in both experiments indicates the occupation of $Mn^{2+}$ ions in the non-perturbed cation sites and the decay component of nearly 1 ms probably belongs to the $Mn^{2+}$ ions occupied in the lattice sites closer to the surface of the nanocrystals. However, the additional low-intensity decay component that shifted from milliseconds to microseconds during the high-pressure study using the pulsed laser is, either from the $Mn^{2+}$ ions occupied on the lattice sites of the surface of the nanocrystal or from the tale of the blue emission that overlapped with the orange band related to the $^4T_1 \rightarrow {}^6A_1$ of $Mn^{2+}$ [49]. Moreover, a similar three exponential decay behavior was also observed in ZnS phosphor doped with 5% $Mn^{2+}$ synthesized via reverse micelle route[50], and the minor size difference in ZnS:$Mn^{2+}$ nanoparticles also significantly influences the decay time values [51].

The $Mn^{2+}$ ion luminescence in ZnS in Figure 2(a) is also interpreted with the crystal field theory and Tanabe Sugano (T-S) diagrams; it explains the electronic absorption energy levels and crystal field strength of the $Mn^{2+}$ present in the material. The strength of the crystal field is also influenced by the size of the nanoparticle. Due to the half-filled $3d^5$ electronic configuration of the $Mn^{2+}$ ion, the $d^5$ configuration of the T-S diagram is used for the calculations and it is designed with Racah parameters B, C, and crystal field strength Dq. The most probable values for this can be obtained by the fit of the crystal-field theory and 10Dq= $\Delta$= 4526.32 cm−1, B= 580.45 cm−1, C=3140.18 cm−1, $\Delta$/B= 7.8, and C/B= 5.41. Moreover theoretically obtained crystal field levels $^4T_1(^4G)$ at 528 nm, $^4T_2(^4G)$ at 482 nm, $^4A_1$, and $^4E$ $(^4G)$ at 465 nm are comparable with the experimentally obtained values in Table1. The Racah parameters B and C were calculated from the following equations:

$$^4A_{1g} + {}^4E_g (^4G) \rightarrow 10B+5C \tag{4}$$

$$^4E_g (^4D) \rightarrow 17B+5C \tag{5}$$

The experimentally obtained values for $^4E_g (^4D)$ and $^4A_{1g} + {}^4E_g (^4G)$ are 391 nm (21505 cm$^{-1}$) and 465 nm (21505 cm$^{-1}$) respectively from the excitation spectra and the direct calculation of equations (4) & (5) shows Racah parameters to be B = 581.5 cm-1, C = 3138 cm-1 and C/B = 5.4. As pressure increases, the main $Mn^{2+}$ luminescence band related to the $^4T_1 \rightarrow {}^6A_1$ transition changes towards the higher crystal field in the T-S diagram, In other words, the energy of this quartet $^4T_{1g}$ level progressively diminishes as the strength of the crystal field increases. Figure 5(c) shows a linear decrease in the energy of the main luminescence band with the increase of pressure but the $Mn^{2+}$ luminescence band is completely quenched before the crossing point between the quartet $^4T_{1g}$ level and the doublet $^2T_2$ level in the T-S diagram due to the zinc blende to rock salt phase transition near 16.47 GPa. However, the strongly spin forbidden transitions to the $^2T_2$ level are not visible in the excitation spectra. Previous pressure-induced quenching of $Mn^{2+}$ luminescence belonging to the same $d^5$ system was observed because of the crossing between the radiative $^4T_{1g}$ level and the non-radiative $^2T_2$ level[52]. Another pressure-induced study of $Fe^{3+}$ doped oxide sample belonging to the same $d^5$ class of system shows a luminescence quenching near 14 GPa due to phase transition [53]. Despite the quenching of the $^4T_1 \rightarrow {^6A_1}$ luminescence band (due to the aforementioned phase transition and a decrease in the distance between the $^4T_{1g}$ level and the non-radiative $^2T_2$ level with the increase of pressure), it is evident that the overall decay times of the $Mn^{2+}$ ions in ZnS are weekly affected by the increase in strength of the crystal field.

## 5. Conclusion

Pressure and temperature-dependent luminescence analysis of zinc blende structured ZnS nanophosphor doped with 2% of $Mn^{2+}$ was successfully performed. The phosphor contains nanocrystallites with near quantum dot size (of 2.38 nm) that are synthesized by the chemical precipitation method, showing intense UV and X-ray induced luminescence at ambient conditions. Also, it has highly desirable self-powered mechano-luminescence properties at ambient temperature conditions without any external irradiation source and it gradually increases with external irradiation. Additionally, the high-pressure luminescence analysis using a diamond anvil cell confirms that the bright orange $Mn^{2+}$ luminescence shifted from visible to the near-infrared spectral region and was completely quenched due to a reversible phase transition from zinc blende to rocksalt structure near 16.41 GPa. The sample retained its initial bright orange luminescence after the complete release of pressure, compared to the irreversible phase transition of ZnS:$Mn^{2+}$ phosphor with a wurtzite structure [40,54]. The milliseconds decay times also confirm the perfect incorporation of $Mn^{2+}$ ions inside the host lattice and show that the multi-exponential $Mn^{2+}$ decay time is influenced by both pressure and temperature.

**Suplementary material:**

The supplementary material contains results of EPR and SQUID measurements of both ZB and WZ samples, showing that the $Mn^{2+}$ concentration is about 19 times larger in WZ than in the ZB sample.

**Acknowledgements**

This work was partially supported by the National Science Centre, Poland, grant SHENG 2 number: 2021/40/Q/ST5/00336 and NCN project number 2019/33/B/ST8/02142.